\documentclass[11pt]{article}
\usepackage{times}
\usepackage{mathfont}
\usepackage{cite}
\usepackage{url}
\def\min{\mathop{\rm min}}

\mathcode`O="724F

\DeclareSymbolFont{lasy}{U}{lasy}{m}{n}
\let\Box\undefined
\DeclareMathSymbol\Box{0}{lasy}{"32}

\newtheorem{lemma}{Lemma}

\newtheorem{theorem}{Theorem}
\newcommand{\qed}{\hfill$\Box$\medbreak}
\newenvironment{proof}{\noindent{\bf Proof: }}{\qed}

\begin{document}

\title{Small Maximal Independent Sets and Faster Exact Graph Coloring}

\author{David Eppstein\thanks{Dept. Inf. \& Comp. Sci., Univ. of
California, Irvine, CA 92697-3425.  Email: {\tt eppstein@ics.uci.edu}.}}

\date{ }
\maketitle

\begin{abstract}
We show that, for any $n$-vertex graph $G$ and integer parameter $k$,
there are at most $3^{4k-n} 4^{n-3k}$
maximal independent sets $I\subset G$ with $|I|\le k$, and that all
such sets can be listed in time
$O(3^{4k-n} 4^{n-3k})$. These bounds are tight when $n/4 \le k
\le n/3$. As a consequence, we show how to compute the exact chromatic
number of a graph in time $O((4/3 + {3^{4/3}}/{4})^n)
\approx 2.4150^n$,
improving a previous
$O((1+3^{1/3})^n)\approx 2.4422^n$ algorithm of Lawler (1976).
\end{abstract} 

\section{Introduction}

One of the earliest works in the area of worst-case analysis of NP-hard
problems is a 1976 paper by Lawler~\cite{Law-IPL-76} on graph
coloring.  It contains two results: an algorithm for finding a
3-coloring of a graph (if the graph is 3-chromatic) in time
$O(3^{n/3})\approx 1.4422^n$, and an algorithm for finding the chromatic
number of an arbitrary graph in time
$O((1+3^{1/3})^n)\approx 2.4422^n$.
Since then, the area has grown, and there has been a sequence of papers
improving Lawler's 3-coloring algorithm
\cite{Sch-WG-93,BeiEpp-FOCS-95,cs.DS/0006046,Epp-SODA-01},
with the most recent algorithm taking time $\approx 1.3289^n$.
However, there has been no improvement to Lawler's chromatic number
algorithm.

Lawler's algorithm follows a simple dynamic programming approach, in
which we compute the chromatic number not just of $G$ but of all its
induced subgraphs.  For each subgraph $S$, the chromatic number is found
by listing all maximal independent subsets $I\subset S$, adding one to the
chromatic number of
$S\setminus I$, and taking the minimum of these values.  The
$O((1+3^{1/3})^n)$ running time of this technique follows
from an upper bound of $3^{n/3}$ on the number of maximal independent
sets in any $n$-vertex graph, due to Moon and
Moser~\cite{MooMos-IJM-65}.  This bound is tight in graphs formed by a
disjoint union of triangles.

In this paper, we provide the first improvement to Lawler's algorithm,
using the following ideas.  First, instead of removing a maximal
independent set from each induced subgraph $S$, and computing the
chromatic number of
$S$ from that of the resulting subset, we add a maximal independent set of
$G\setminus S$ and compute the chromatic number of the resulting
superset from that of $S$.  This reversal does not itself affect the
running time of the dynamic programming algorithm, but it allows us to
constrain the size of the maximal independent sets we consider to
at most $|S|/3$. We show that, with such a constraint, we can improve the
Moon-Moser bound: for any $n$-vertex graph $G$ and integer parameter $k$,
there are at most $3^{4k-n} 4^{n-3k}$
maximal independent sets $I\subset G$ with $|I|\le k$.
This bound then leads to a corresponding improvement in the
running time of our chromatic number algorithm.

\section{Preliminaries}

We assume as given a graph $G$ with vertex set $V(G)$ and edge set
$E(G)$.  We let $n=|V(G)|$ and $m=|E(G)|$.  A {\em proper coloring} of
$G$ is an assignment of colors to vertices such that no two endpoints of
any edge share the same color.  We denote the chromatic number of $G$ (the
minimum number of colors in any proper coloring) by $\chi(G)$.

\def\deg{\mathop{\rm deg}}

If $V(G)=\{v_0,v_1,\ldots v_{n-1}\}$, then we can place subsets
$S\subseteq V(G)$ in one-to-one correspondence with the integers
$0,1,\ldots 2^n-1$:
$$S\leftrightarrow \sum_{v_i\in S} 2^i.$$
Subsets of vertices also correspond to induced subgraphs of $G$,
in which we include all edges between vertices in the subset.
We make no distinction between these three equivalent views of a
vertex subset, so e.g. we will write $\chi(S)$ to indicate the chromatic
number of the subgraph induced by set $S$, and $X[S]$ to indicate a
reference to an array element indexed by the number $\sum_{v_i\in S} 2^i$.
We write $S<T$ to indicate the usual arithmetic comparison between two
numbers, and $S\subset T$ to indicate the usual (proper) subset relation
between two sets.  Note that, if $S\subset T$, then also $S< T$, although
the reverse implication does not hold.

A set $S$ is a {\em maximal $k$-chromatic subset} of $T$
if $S\subseteq T$, $\chi(S)=k$, and $\chi(S')>k$ for every $S\subset
S'\subseteq T$. In particular, if $k=1$, $S$ is a {\em maximal independent
subset} of $T$.

For any vertex $v\in V(G)$, we let $N(v)$ denote the set of neighbors of
$v$, including $v$ itself.  If $S$ and $T$ are sets, $S\setminus T$
denotes the set-theoretic difference, consisting of elements of $S$ that
are not also in $T$.  $K_i$ denotes the complete graph on $i$ vertices.
We write $\deg(v,S)$ to denote the degree of vertex
$v$ in the subgraph induced by $S$.

We express our pseudocode in a syntax similar to that of C, C++, or Java.
In particular this implies that array indexing is zero-based.
We assume the usual RAM model of computation, in which a single
cell is capable of storing an integer large enough to index the memory
requirements of the program (thus, in our case, $n$-bit values are
machine integers), and in which arithmetic and array indexing operations
on these values are assumed to take constant time.

\section{Small Maximal Independent Sets}

\begin{theorem}
\label{thm:smallmis}
Let $G$ be an $n$-vertex graph, and $k$ be a nonnegative number.
Then the number of maximal independent sets $I\subset V(G)$
for which $|I|\le k$ is at most $3^{4k-n} 4^{n-3k}$.
\end{theorem}

\begin{proof}
We use induction on $n$; in the base case $n=0$, there is one (empty)
maximal independent set, and for any $k\ge 0$, $1\le
3^{4k}4^{-3k}=(81/64)^k$. Otherwise, we divide into cases according to the
degrees of the vertices in $G$, as follows:

\begin{itemize}
\item
If $G$ contains a vertex $v$ of degree three or more,
then each maximal independent set $I$ either contains $v$
(in which case $I\setminus\{v\}$ is a maximal independent set of
$G\setminus N(v)$) or it does not contain $v$
(in which case $I$ itself is a maximal independent set of
$G\setminus\{v\}$).  Thus, by induction, the number of maximal
independent sets of cardinality at most $k$ is at most
$$
3^{4k-(n-1)} 4^{(n-1)-3k} + 3^{4(k-1)-(n-4)} 4^{(n-4)-3(k-1)}
= (\frac34 + \frac14) 3^{4k-n} 4^{n-3k} = 3^{4k-n} 4^{n-3k}$$
as was to be proved.

\item
If G contains a degree-one vertex $v$,
let its neighbor be $u$.
Then each maximal independent set contains exactly one of $u$ or $v$,
and removing this vertex from the set produces a maximal independent
set of either $G\setminus N(u)$ or  $G\setminus N(v)$.
If the degree of $u$ is $d$, this gives us by induction a bound of
$$3^{4(k-1) - (n-2)} 4^{(n-2)-3(k-1)}
+3^{4(k-1) - (n-d-1)} 4^{(n-d-1)-3(k-1)}
\le \frac89\, 3^{4k-n} 4^{n-3k}$$
on the number of maximal independent sets of cardinality at most $k$.

\item If $G$ contains an isolated vertex $v$,
then each maximal independent set contains $v$,
and the number of maximal independent sets of cardinality at most $k$
 is at most
$$3^{4(k-1)-(n-1)} 4^{(n-1)-3(k-1)}
=\frac{16}{27}\, 3^{4k-n} 4^{n-3k}.$$

\item If $G$ contains a chain $u$-$v$-$w$-$x$
of degree two vertices,
then each maximal independent set contains $u$,
contains $v$, or does not contain $u$ and contains $w$.
Thus in this case the number of maximal independent sets
of cardinality at most $k$
is at most
$$2 \cdot 3^{4(k-1) - (n-3)} 4^{(n-3)-3(k-1)}
+ 3^{4(k-1) - (n-4)} 4^{(n-4)-3(k-1)}
= \frac{11}{12}\, 3^{4k-n} 4^{n-3k}.$$

\item In the remaining case, $G$ consists of a disjoint union of
triangles, all maximal independent sets have exactly $n/3$ vertices,
and there are exactly $3^{n/3}$ maximal independent sets.
If $k\ge n/3$, then $3^{n/3}\le 3^{4k-n} 4^{n-3k}.$
If $k<n/3$, there are no maximal independent sets
of cardinality at most $k$.
\end{itemize}

Thus in all cases the number of maximal independent sets is within the
claimed bound.
\end{proof}

Croitoru~\cite{Cor-COR-79} proved a similar bound with the stronger
assumption that all maximal independent sets have $|I|\le k$.
When $n/4\le k\le n/3$, our result is tight, as can be seen for
a graph formed by the disjoint union of $4k-n$ triangles and $n-3k$
$K_4$'s.

\begin{figure}[p]
\begin{tabbing}
// List maximal independent subsets of $S$ smaller than a given
parameter. \\
// \qquad $S$ is a set of vertices forming an induced subgraph in $G$,\\
// \qquad $I$ is a set of vertices to be included in the MIS (initially
zero), and\\
// \qquad $k$ bounds the number of vertices of $S$ to add to $I$. \\
// We call processMIS$(I)$ on each generated set.  Some non-maximal sets
may be \\
// generated along with the maximal ones, but all generated sets are
independent. \\
\\
void smallMIS (set $S$, set $I$, int $k$) \\
\{ \quad \= \\
   \> if $(S = 0$ or $k=0)$ processMIS(I); \\
   \> else if $($there exists $v\in S$ with $\deg(v,S)\ge 3)$ \\
   \> \{ \quad \= \\
   \> \> smallMIS ($S\setminus\{v\}$, $I$, $k$); \\
   \> \> smallMIS ($S\setminus N(v)$, $I\cup\{v\}$, $k-1$); \\
   \> \} \\
   \> else if $($there exists $v\in S$ with $\deg(v,S) = 1)$ \\
   \> \{ \quad \= \\
   \> \> let $u$ be the neighbor of $v$; \\
   \> \> smallMIS ($S\setminus N(u)$, $I\cup\{u\}$, $k-1$); \\
   \> \> smallMIS ($S\setminus N(v)$, $I\cup\{v\}$, $k-1$); \\
   \> \} \\
   \> else if $($there exists $v\in S$ with $\deg(v,S) = 0)$ \\
   \> \> smallMIS ($S\setminus\{v\}$, $I\cup\{v\}$, $k-1$); \\
   \> else if $($some cycle in $S$ is not a triangle or $k\ge |S|/3)$\\
   \> \{ \\
   \> \> let $u$, $v$, and $w$ be adjacent degree-two
vertices, such that (if possible) $u$ and $w$ are nonadjacent; \\
   \> \> smallMIS ($S\setminus N(u)$, $I\cup\{u\}$, $k-1$); \\
   \> \> smallMIS ($S\setminus N(v)$, $I\cup\{v\}$, $k-1$); \\
   \> \> smallMIS ($S\setminus (\{u\}\cup N(w))$, $I\cup\{w\}$, $k-1$); \\
   \> \} \\
\}
\end{tabbing}
\caption{Algorithm for listing all small maximal independent sets.}
\label{fig:misalg}
\end{figure}

\begin{theorem}
\label{thm:misalg}
There is an algorithm for listing all maximal independent sets smaller
than $k$ in an $n$-vertex graph $G$, in time $O(3^{4k-n} 4^{n-3k})$.
\end{theorem}

\begin{proof}
We use a recursive backtracking search, following the case analysis
of Theorem~\ref{thm:smallmis}: if there is a high-degree vertex, we try
including it or not including it; if there is a degree-one vertex, we
try including it or its neighbor; if there is a degree-zero vertex, we
include it; and if all vertices form chains of degree-two vertices, we
test whether the parameter $k$ allows any small maximal independent
sets, and if so we try including each of a chain of three adjacent
vertices. The same case analysis shows that this algorithm performs
 $O(3^{4k-n} 4^{n-3k})$ recursive calls.

Each recursive call can easily be
implemented in time polynomial in the size of the graph passed to the
recursive call.  Since
our $3^{4k-n} 4^{n-3k}$ bound is exponential in $n$, even when $k=0$,
this polynomial overhead at the higher levels of the recursion is
swamped by the time spent at lower levels of the recursion,
and does not appear in our overall time bound.
\end{proof}

A more detailed pseudocode description of the algorithm is shown in
Figure~\ref{fig:misalg}.  The given pseudocode may generate non-maximal
as well as maximal independent sets, because (when we try not including
a high degree vertex) we do not make sure that a neighbor is later
included.  This will not cause problems for our chromatic number
algorithm, but if only maximal independent sets are desired one can
easily test the generated sets and eliminate the non-maximal ones.
The pseudocode also omits the data structures necessary to implement
each recursive call in time polynomial in $|S|$ instead of polynomial
in the number of vertices of the original graph.

\section{Chromatic Number}

\begin{figure}[t]
\begin{tabbing}
int chromaticNumber (graph $G$) \\
\{ \quad \= \\
   \> int $X[2^n]$; \\
   \> for $(S = 0$; $S\le 2^n$; $S$++) \\
   \> \{ \quad \= \\
   \> \> if $(\chi(S)\le 3)$ $X[S] = \chi[S]$; \\
   \> \> else $X[S] = \infty$; \\
   \> \} \\
   \> for $(S = 0$; $S\le 2^n$; $S$++) \\
   \> \{ \quad \= \\
   \> \> if $(3\le X[S] < \infty)$ \\
   \> \> \{ \quad \= \\
   \> \> \qquad for (each maximal independent set $I$ of $G\setminus S$
with
$|I|\le \displaystyle\frac{|S|}{X[S]}$)\\
   \> \> \qquad \qquad $X[S\cup I] = \min (X[S\cup I], X[S]+1)$;\\
   \> \> \} \\
   \> \} \\
   \> return $X[V(G)]$; \\
\}
\end{tabbing}
\caption{Algorithm for computing the chromatic number of a graph.}
\label{fig:cnalg}
\end{figure}

We are now ready to describe our algorithm for computing the chromatic
number of graph $G$.  We use an array $X$, indexed by the $2^n$ subsets
of $G$, which will (eventually) hold the chromatic numbers of certain of
the subsets including $V(G)$ itself.  We initialize
this array by testing, for each subset
$S$, whether $\chi(S)\le 3$; if so, we set $X[S]$ to $\chi(S)$, but
otherwise we set $X[S]$ to $\infty$.

Next, we loop through the subsets $S$ of $V(G)$, in numerical order
(or any other order such that all proper subsets of each set $S$ are
visited before we visit $S$ itself).
When we visit $S$, we first test whether $X[S]\ge 3$.  If not, we skip
over $S$ without doing anything.  But if $X[S]\ge 3$, we loop through
the small independent sets of $G\setminus S$, limiting the size of each
such set to $|S|/X[S]$, using the algorithm of the previous section.
For each independent set $I$, we set $X[S\cup I]$ to the minimum of
its previous value and $X[S]+1$.

Finally, after looping through all subsets, we return the value in
$X[V(G)]$ as the chromatic number of $G$.
Pseudocode for this algorithm is shown in Figure~\ref{fig:cnalg}.

\begin{lemma}
Throughout the course of the algorithm, for any set $S$,
$X[S]\ge\chi(S)$.
\end{lemma}

\begin{proof}
Clearly this is true of the initial values of $X$.
Then for any $S$ and any independent set $I$,
we can color $S\cup I$ by using a coloring of $S$
and another color for each vertex in $I$, so
$\chi(S\cup I)\le \chi(S)+1\le X[S]+1$,
and each step of our algorithm preserves the invariant. 
\end{proof}

\begin{lemma}\label{lem:MSI}
Let $M$ be a maximal $k+1$-chromatic subset of $G$,
and let $(S,I)$ be a partition of $M$ into a $k$-chromatic subset $S$
and an independent subset $I$, maximizing the cardinality of $S$ among
all such partitions. Then
$I$ is a maximal independent subset
of $G\setminus S$ with $|I|\le |S|/k$,
and $S$ is a maximal $k$-chromatic subset of $G$.
\end{lemma}

\begin{proof}
If we have any $(k+1)$-coloring of $G$,
then the partition formed by separating the largest $k$ color classes
from the smallest color class satisfies the inequality
$|I|\le |S|/k$, so clearly this also is true when $(S,I)$ is the partition
maximizing $|S|$.  If $I$ were not maximal, due to the existence
of another independent set $I\subset I'\subset G\setminus S$,
then $S\cup I'$ would be a larger $(k+1)$-chromatic graph, violating the
assumption of maximality of $M$.

Similarly, suppose there were another $k$-chromatic set
$S\subset S'\subset G$.
Then if $S'\cap I$ were empty, $S'\cup I$ would be a $(k+1)$-chromatic
superset of $M$, violating the assumption of $M$'s maximality.
But if $S'\cap I$ were nonempty, $(S',I\setminus S')$
would be a better partition than $(S,I)$,
so in either case we get a contradiction.
\end{proof}

\begin{lemma}\label{lem:cnok}
Let $M$ be a maximal $k+1$-chromatic subset of $G$.
Then, when the outer loop of our algorithm reaches $M$,
it will be the case that $X[M]=\chi(M)$.
\end{lemma}

\begin{proof}
Clearly, the initialization phase of the algorithm causes
this to be true when $\chi(M)\le 3$. Otherwise,
let $(S,I)$ be as in Lemma~\ref{lem:MSI}.
By induction on $|M|$, $X[S]=\chi(S)$
at the time we visit $S$.
Then $X[S]\ge 3$, and $|I|\le |S|/X[S]$,
so the inner loop for $S$ will visit $I$ and
set $X[M]$ to $X[S]+1=\chi(M)$.
\end{proof}

\begin{theorem}\label{thm:cnalg}
We can compute the chromatic number of a graph $G$
in time $O((4/3 + {3^{4/3}}/{4})^n)$ and space $O(2^n)$.
\end{theorem}

\begin{proof}
$V(G)$ is itself a maximal $\chi(G)$-chromatic subset of $G$,
so Lemma~\ref{lem:cnok} shows that the algorithm correctly computes
$\chi(G)=X[V(G)]$.  Clearly, the space is bounded by $O(2^n)$. It remains
to analyze the algorithm's time complexity.

First, we consider the time spent initializing $X$.
Since we perform a 3-coloring algorithm on each subset of $G$,
this time is
$$\sum_{S\subset V(G)} O(1.3289^{|S|})
= O\Bigl(\sum_{i=0}^n {n\choose i} 1.3289^i\Bigr)
= O(2.3289^i).$$

Finally, we bound the time in the main loop of the algorithm.
We may possibly apply the algorithm of Theorem~\ref{thm:misalg}
to generate small independent subsets of
each set $G\setminus S$.  In the worst case, $X[S]=3$
and we can only limit the size of the generated independent sets
to $|S|/3$.  We spend constant time adjusting the value of
$X[S\cup I]$ for each generated set. Thus, the total time can be bounded
as
$$\sum_{S\subset V(G)} O(3^{4\frac{|S|}{3}-|G\setminus S|}
4^{|G\setminus S|-3\frac{|S|}{3}})
= O\Bigl(\sum_{i=0}^n {n\choose i} 3^{\frac{7i}{3}-n} 4^{n-2i}\Bigr)
= O\Bigl( (\frac43 + \frac{3^{4/3}}{4})^n \Bigr).$$
This final term dominates the overall time bound.
\end{proof}

\section{Finding a Coloring}

\begin{figure}[t]
\begin{tabbing}
void color (graph $G$) \\
\{ \quad \= \\
   \> compute array $X$ as in Figure~\ref{fig:cnalg};\\
   \> $S = V(G)$; \\
   \> for $(T = 2^n-1$; $T\ge 0$; $T${}$-${}$-)$ \\
   \> \{ \quad \= \\
   \> \> if $(T\subset S$ and $X[S\setminus T]=1$ and $X[T]=X[S]-1$)\\
   \> \> \{ \quad \= \\
   \> \> \> color all vertices in $S\setminus T$ with the same new
color;\\
   \> \> \> $S = T$;\\
   \> \> \} \\
   \> \} \\
\}
\end{tabbing}
\caption{Algorithm for optimally coloring a graph.}
\label{fig:color}
\end{figure}

Although the algorithm of the previous section finds the chromatic
number of $G$, it is likely that an explicit coloring is desired,
rather than just this number.
The usual method of performing this sort of construction task in a
dynamic programming algorithm is to augment the dynamic programming
array with back pointers indicating the origin of each value computed in
the array, but since storing $2^n$ chromatic numbers is likely to be the
limiting factor in determining how large a graph this algorithm can be
applied to, it is likely that also storing $2^n$ set indices will
severely reduce its applicability.

Instead, we can simply search backwards from $V(G)$
until we find a subset $S$ that can be augmented by an independent set
to form $V(G)$, and that has chromatic number $\chi(S)=\chi(G)-1$
as indicated by the value of $X[S]$.
We assign the first color to $G\setminus S$.
Then, we continue searching for a similar subset $T\subset S$, etc., until
we reach the empty set.
Although not every set $S$ may necessarily have $X[S]=\chi(S)$,
it is guaranteed that for any $S$
we can find $T\subset S$ with $S\setminus T$ independent and
$X[T]=X[S]-1$, so this search procedure always finds a correct coloring.

\begin{theorem}
After computing the array $X$ as in Theorem~\ref{thm:cnalg},
we can compute an optimal coloring of $G$
in additional time $O(2^n)$ and no additional space.
\end{theorem}

We omit the details of the correctness proof and analysis.
A pseudocode description of the coloring algorithm
is shown in Figure~\ref{fig:color}.

\section{Conclusions}

We have shown a bound on the number of small independent sets in a
graph, shown how to list all small independent sets in time proportional
to our bound, and used this algorithm in a new dynamic programming
algorithm for computing the chromatic number of a graph as well as an
optimal coloring of the graph.

Our bound on the number of small independent sets is tight for $n/4\le
k\le n/3$, and an examination of the analysis of Theorem~\ref{thm:cnalg}
shows that independent set sizes in this range are also the ones leading
to the worst case time bound for our coloring algorithm.  Nevertheless, it
would be of interest to find tight bounds on the number of small
independent sets for all ranges of $k$.  It would also be of interest
to find an algorithm for
listing all small maximal independent sets in time proportional to the
number of generated sets rather than simply proportional to the worst case
bound on this number.

Our worst case analysis of the chromatic number algorithm assumes that,
every time we call the procedure for listing small maximal independent
sets, this procedure achieves its worst case time bound.  But is it
really possible for all sets $G\setminus S$ to be worst case instances
for this procedure? If not, perhaps the analysis of our
coloring algorithm can be improved.

Can we prove a bound smaller than $n\choose i$ on the number of
$i$-vertex maximal $k$-chromatic induced subgraphs of a graph $G$?
If such a bound could be proven, even for $k=3$, we could likely improve
the algorithm presented here by only looping through the independent
subgraphs of $G\setminus S$ when $S$ is maximal.

An alternative possibility for improving the present algorithm would
be to find an algorithm for testing whether $\chi(G)\le 4$ in time
$o(1.415^n)$. Then we could test the four-colorability of all subsets of
$G$ before applying the rest of our algorithm, and avoid looping over
maximal independent subsets of $G\setminus S$ unless $X[S]\ge 4$.  This
would produce tighter limits on the independent set sizes and therefore
reduce the number of independent sets examined.  However such a time
bound would be significantly better than what is currently known for
four-coloring algorithms.

\bibliographystyle{abuser}
\bibliography{3color}

\end{document}